\documentclass[prl,aps,amssymb,amsmath,superscriptaddress,twocolumn,showpacs,]{revtex4-1}

\usepackage{graphicx}

\renewcommand{\epsilon}{\varepsilon}

\renewcommand{\phi}{\varphi}

\def\be{\begin{equation}}
\def\ee{\end{equation}}
\def\bea{\begin{eqnarray}}
\def\eea{\end{eqnarray}}

\def\xx{{\bf x}}
\def\yy{{\bf y}}

\let\s=\sigma

\begin{document}

\title{Realization of stripes and slabs in two and three dimensions}

\author{Alessandro Giuliani}
\affiliation{Dipartimento di Matematica e Fisica, Universit\`a di Roma
Tre, L.go S. L. Murialdo 1, 00146 Roma, Italy}

\author{Elliott H. Lieb}
\affiliation{Department of Mathematics,
Princeton University, Washington Road, Princeton, NJ 08544, USA}
\affiliation{Department of Physics,
Princeton University, P.~O.~Box 708, Princeton, NJ 08542, USA}

\author{Robert Seiringer}
\affiliation{Institute of Science and Technology Austria (IST Austria), Am
Campus 1, A - 3400 Klosterneuburg, Austria}

\begin{abstract}
We consider Ising models in two and three dimensions with nearest neighbor
ferromagnetic interactions and long range, power law decaying,
antiferromagnetic
interactions. If the strength of the ferromagnetic coupling $J$ is larger 
than a critical value $J_c$, then the ground state is homogeneous and
ferromagnetic. As the critical value is approached from smaller values of
$J$, it is believed that the ground state consists of a periodic array of
stripes ($d=2$) or slabs ($d=3$), all of the
same size and alternating magnetization. Here we prove 
rigorously that the ground state energy per site converges to that 
of the optimal periodic striped/slabbed state, in the limit that $J$ tends
to the
ferromagnetic transition point. While this theorem does not prove rigorously
that the ground state is precisely striped/slabbed, it does prove that in
any
suitably large box the ground state is striped/slabbed with high
probability.
\end{abstract}


\maketitle


The spontaneous emergence of periodic states in translation invariant 
systems is still an incompletely understood phenomenon. Even less well
understood is the phenomenon of translation symmetry breaking 
in only one direction, that is to say formation of striped patterns in two
dimensions or slabbed patterns in three dimensions, which we collectively
refer to as stripes.
Particularly interesting is the formation of wide stripes, by which 
we mean that stripes have a width much larger than the microscopic length scales. Stripes of this kind are expected to display a sort
of universal 
phenomenology, which is in fact observed in a variety of different systems, 
ranging from magnetic films \cite{CMST06,DMW00,MWRD95,SW92,SS99}, to manganites \cite{SJ01}, to high-temperature superconductors \cite{EK93, KKACVFM10, KBFOTKH03,  RRB04, TSANU95}, MOSFETs \cite{S03,SK04},
polymer suspensions \cite{HACSSHRC00, MP04}, 
twinned martensites \cite{KM94,GM12}, Coulomb glasses \cite{GTV00}, and many others \cite{BS09,CPPV09,CDSN12,CN11,EJ10,MP03,OLD08,VSPPP08}.

\begin{figure}[h]
\centering
\includegraphics[width=.38\textwidth]{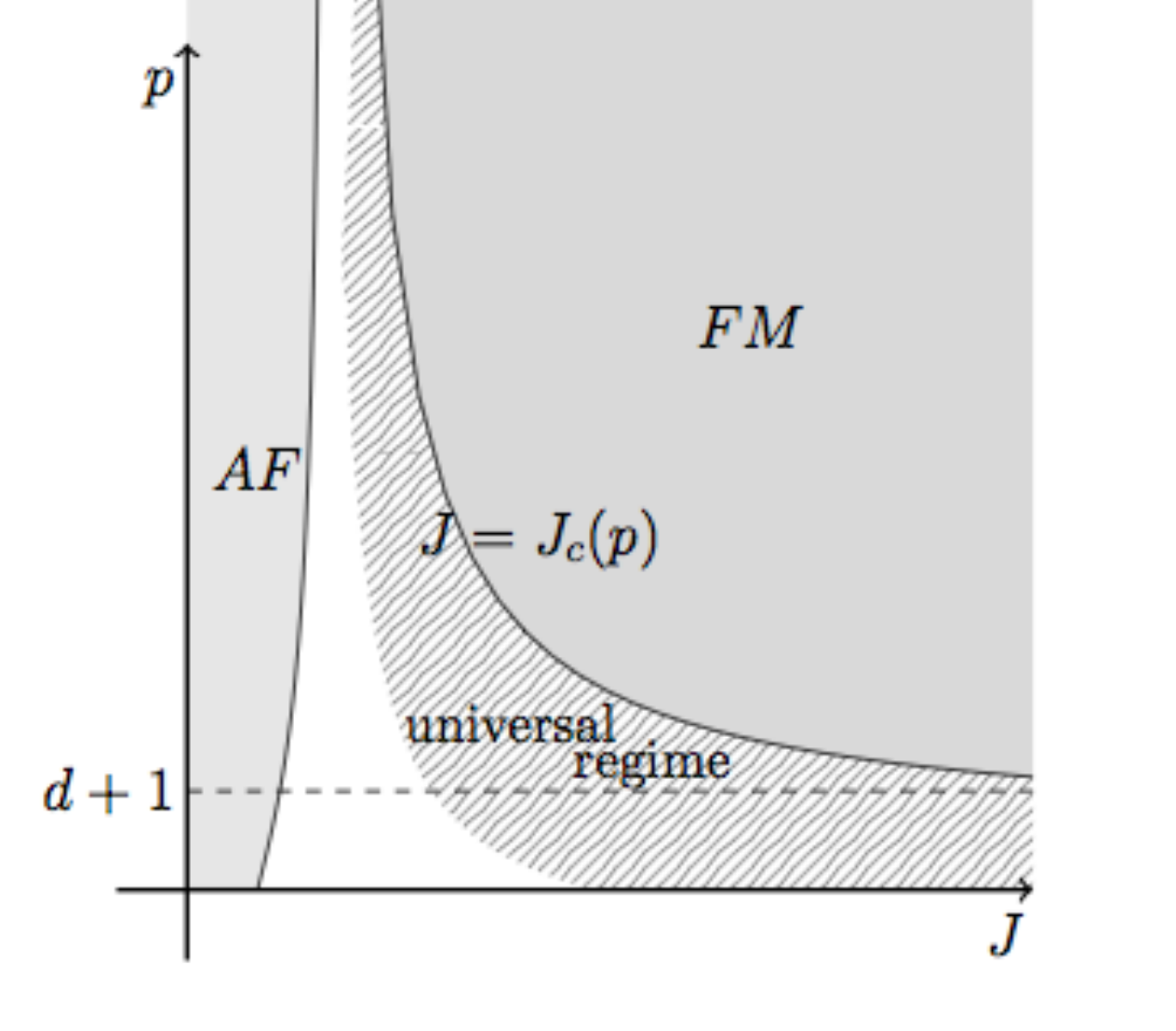}
\caption{Ground state phase diagram as a function of the
antiferromagnetic decay exponent $p$ and of the ferromagnetic strength $J$. 
In the leftmost region the ground state is the N\'eel antiferromagnetic
state, while in the rightmost region it is the homogeneous
ferromagnetic state. These two phases are rigorously known, while the rest
of the phase diagram remains to be understood. The light-gray shaded region is the
``universal regime'', where the ferromagnetic islands (droplets) have
typical size much larger than the lattice spacing. The conjecture is that
the ground state is periodic and striped in the whole universal regime. A
partial proof of this fact is given in this paper.
}
\label{fig1}
\end{figure}

While there exist some rigorous examples of symmetry breaking in two
dimensions into doubly-periodic crystalline structures
\cite{Kennedy-Lieb,Radin,Suto,Theil}, we are aware of only one rigorous proof 
of formation of periodic arrays of wide stripes 
in isotropic two-dimensional systems: this is a system of in-plane 
spins with four possible orientations interacting via a short range
exchange plus the actual three-dimensional dipolar interaction \cite{GLL07}.
It would be nice to find more examples of this kind. A simple and very
popular model used to understand stripe formation in the classical setting 
is a $d$-dimensional Ising model with the following Hamiltonian:
\be
H=-J\sum_{\langle
\xx,\yy\rangle}(\s_\xx\s_\yy-1)+\sum_{\{\xx,\yy\}}\frac{(\s_\xx\s_\yy-1)}{
|\xx-\yy|^p}\label{1}
\ee
where $J>0$ is the relative strength of the attractive exchange
interaction, the first sum ranges over nearest neighbor pairs in $\mathbb
Z^d$, $d=2,3$, and the second over pairs of distinct sites in $\mathbb Z^d$.
Depending on the specific value of the exponent $p$, the 
second term in the Hamiltonian can describe a Coulomb ($p=1$), a dipolar
($p=3$), or a more general repulsive interaction. Note that the
Hamiltonian is normalized so that the homogeneous ferromagnetic state has
zero energy. The question is to determine the ground state of the system,
as the parameters $J$ and $p$ are varied. In some limiting cases, it
is
easy to identify the minimal energy states: e.g., if $J$ is
sufficiently small, the ground state is the N\'eel antiferromagnet, as one
can prove by using reflection positivity \cite{FILS2}. If $p>d+1$, there
exists a critical value $J_c(p)$, of the form 
$$J_c(p)=\sum_{y_1=1}^\infty\ \sum_{y_2,\dots,y_d=-\infty}^\infty\frac{y_1}{
(y_1^2+\cdots+y_d^2)^ {
p/2 } }\;, $$
such that the
homogeneous ferromagnetic state is the ground state for $J\ge J_c$, and it
is not the ground state for $J<J_c$ \cite{GLL11}. Note that $J=J_c(p)$ is
the value of the ferromagnetic strength at which the surface tension of an
infinite, isolated, straight domain wall vanishes. The expected region where
wide stripes should occur is $p\le d+1$ if $J\gg 1$, and $p>d+1$ if
$J\lesssim J_c$. This is the region that we call ``universal'',
in the sense that the structures displayed by the ground state in this
regime are large compared to the lattice spacing and, therefore,
their shape is expected to be independent of the microscopic details of the
Hamiltonian. See Fig.\ref{fig1}.

In this article, we report a recent advance
in the understanding of the ground state phase diagram of model (\ref{1})
in the universal regime, for $p>2d$. Our new estimates are the
sharpest rigorous bounds available on the ground state of the 
class of models under consideration here. To the best of our knowledge,
these are the first results of this kind in three dimensions. 
Before we state them, let us
introduce a few more definitions. Let $e_{\rm s}(h)$ be the energy per site
in the thermodynamic limit of periodic striped configurations consisting of
stripes all of width $h$. We denote by $h(J)$ the optimal stripe width,
which can be obtained by minimizing $e_{\rm s}(h)$ over $h\in\mathbb N$. For
$p>d+1$, $h(J)$ turns out to
be of the order $(J_c-J)^{-\frac1{p-d-1}}$ as $J\nearrow J_c$. Let 
us denote by $e_{\rm S}(J)\equiv e_{\rm s}(h(J))$ the optimal striped
energy per site and by $e_0(J)$ the actual ground state energy per site in
the thermodynamic limit. Note that $e_0(J)=0$ for $J\geq J_c$.
Our main results can be summarized as follows. 
\vskip.2truecm

{\bf Theorem.} {\it Let us consider model (\ref{1}) with $d=2,3$ and $p>2d$. 
As $J\to J_c$ from below, 
$$ \lim_{J\to J_c}\frac{e_0(J)}{e_{\rm S}(J)}=1\;.$$}
\vskip.2truecm

A few remarks are in order. The theorem says that asymptotically, as we
approach the ferromagnetic transition line $J=J_c(p)$, the actual ground
state energy approaches the optimal striped energy,
which is a very strong indication of the conjectured
periodic striped structure of the ground state. The proof comes with
explicit error bounds on the difference $e_0(J)/e_{\rm S}(J)-1$,
namely
\be 1\ge \frac{e_0(J)}{e_{\rm S}(J)}\ge
1+O\big((J_c-J)^{\frac{p-2d}{(d-1)(p-d-1)}}\big)\;.\label{1.1bis}\ee
More precisely, the proof shows that the density of corners in the
minimizing configuration is much smaller than $(J_c-J)^{d/(d-1)}$,
i.e., the
average mutual distance between corners is much larger than the typical
stripe width $h(J)$. By corners here we mean the points ($d=2$) or
edges ($d=3$) where domain walls
bend by 90$^o$. The
notion of corner and corner energy was introduced 
in \cite{GLL11} and understood there to play an important role for the case
$p>2d$: if widely separated from each other, the corners give a finite, positive 
contribution to the energy and, therefore, can be thought of as the
elementary excitation of the system, at least in some approximate sense. 
Our new estimate (\ref{1.1bis}) implies that the ground state
configuration, if restricted to a suitable large window of side $\ell\gg
h(J)$, with high probability has no corners, i.e., with high probability it
is exactly striped. Similarly, we can show that with high probability 
these stripes have width all very close to $h(J)$. 

The proof of the theorem is based on refined lower bounds on the ground
state energy. The details of the proof are lengthy and will be given
elsewhere \cite{GLS}. Here we explain the main strategy behind the proof. These
ideas may prove useful for subsequent developments in this subject. The key
steps are the following.
\begin{enumerate}
\item Representation of the energy in terms of droplets: these are 
simply the maximal connected regions of negative spins, whose boundaries are
the standard low-temperature contours of the nearest neighbor Ising model.
The energy can be written as a sum of droplet self-energies, plus a
long-range antiferromagnetic repulsion among different droplets. 
\item Localization of the droplet energy functional into boxes $Q$ of proper
size, to be optimized over. By localization we mean that we bound from
below the original energy of a generic droplet configuration in terms of 
a sum of independent local energies, each depending only on
the restriction of the droplet configuration to the given box $Q$. Of
course, the non-trivial aspect of this localization bound is due to the long
range nature of the antiferromagnetic potential. The important fact is that
our lower bound is sharp for striped configurations, up to unimportant
boundary corrections. On top of that, we show that the
localized energy of any droplet with one or more corners is positive,
irrespective of any details of the configuration: therefore, 
for the purpose of a lower bound, corners can be eliminated in every box.
\item Minimization of the corner-free configurations by the method
of block reflection positivity, introduced in \cite{GLL06} and further developed in \cite{GLL07,GLL09,GLL09b,BEGM12}. Once the 
corners have been eliminated, we are left with purely striped
configurations,
whose energy can be further bounded from below by iterative reflections
across the straight domain walls. After repeated reflections, we end up
with periodic striped configurations, and the proof is complete. 
\end{enumerate}

We believe that the ground 
state displays striped order also for values $p\le 2d$. However, our proof 
only works for $p>2d$, the reason being two-fold: (i) the energy of an isolated corner 
($d=2$) or of a trihedral vertex ($d=3$)  
becomes infinite at smaller values of $p$ and, therefore, there is no obvious way of 
identifying the local excitations of the system; (ii) the optimal striped energy per site 
is of the same order as that of other putative ordered ground states, such as checkerboard
or columnar states, and, therefore, it is difficult to exclude the emergence of other 
ordered structures on the basis of local energy estimates.

In {\it conclusion}, we considered Ising models in two and three dimensions with nearest neighbor ferromagnetic and power 
law decaying antiferromagnetic interactions. We presented new rigorous bounds on the ground state energy and, in particular, we showed that the actual ground state energy per site tends to the one of the optimal periodic striped configuration, as we approach the ferromagnetic transition line. Moreover, we proved that the minimizing spin configurations are striped in a suitable sense; namely, if restricted to finite windows of proper size (much larger than 
the optimal stripe width), they all look precisely striped with very high, explicitly estimated, probability. 
These are the most refined rigorous bounds on the ground state energy of the considered model, and the first 
of this kind in three dimensions. 
Our new methods, which the proof of the theorem is based on, combine for the first time 
the ideas of energy localization into boxes and of block reflection positivity, in the context of 
isotropic systems with competing interactions in two and three dimensions. We expect them to be 
crucial for further developments in the subject and, in particular, for a proof of exact, macroscopic, stripe ordering
in two and three dimensions. 
\vskip.2truecm

{\bf Acknowledgments.} The research leading to these results has
received funding from the European Research Council under the EU's Seventh Framework Programme ERC Starting Grant CoMBoS (grant
agreement n$^o$ 239694; A.G. and R.S.), the U.S. National Science Foundation
(grant PHY 0965859; E.H.L.), the Simons Foundation (grant \# 230207;
E.H.L.) and the NSERC (R.S.).  The work is part of a project started in
collaboration with Joel Lebowitz, whom we thank for many useful
discussions and for his constant encouragement.


\bibliographystyle{amsalpha}

\end{document}